\documentclass[12pt,aps,prb,preprint,showpacs,showkeys]{revtex4}  

\usepackage{amsmath}    
\usepackage{amsfonts}   
\usepackage{amssymb}
\usepackage{graphicx}   
\usepackage{subfigure}

\begin{document}
\baselineskip = 7.5mm \topsep=1mm
\begin{center}{\LARGE\bf Monte Carlo investigation of the tricritical point stability in
a three-dimensional Ising metamagnet}
\vspace{5 mm}\\ M.\v{Z}ukovi\v{c}$^{\rm *}$ and T.Idogaki
\vspace{5 mm}\\
\end{center}
Department of Applied Quantum Physics, Graduate School of Engineering,
\newline
Kyushu University, Fukuoka 812-8581, Japan \vspace{20 mm}
\newline
\noindent {\bf{Abstract.}} We use Monte Carlo simulations to study multicritical properties of an
Ising metamagnet in an external field. According to the mean field theory predictions, a
three-dimensional layered metamagnet is expected to display a tricritical point decomposition to a
critical endpoint and a bicritical endpoint, when a ratio between intralayer ferromagnetic and
interlayer antiferromagnetic couplings becomes sufficiently small. Our simulations show no evidence of
such a decomposition and produce a tricritical behaviour even for a coupling ratio as small as
$R=0.01$. \vspace{20 mm}
\\ $PACS\ codes$: 75.10.Hk; 75.30.Kz; 75.40.Cx; 75.40.Mg; 75.50.Ee.
\newline
$Keywords$: Ising metamagnet; Monte Carlo simulation; Multicritical behaviour; Tricritical point
decomposition.\\ \vspace{20 mm}
\\ $*$Corresponding author.
\newline
Permanent address: Department of Applied Quantum Physics, Graduate School of Engineering, Kyushu
University, Fukuoka 812-8581, Japan \\ Tel.: +81-92-642-3811; Fax: +81-92-633-6958
\newpage
\noindent {\bf\Large{1.Introduction}} \vspace{3mm}
\newline
\indent Ising metamagnets, systems with ferromagnetic and antiferromagnetic couplings simultaneously
present, have attracted much interest because it is possible to induce novel kinds of critical
behaviour by forcing competition between these couplings, in particular by applying a magnetic field.
They are generally believed to exhibit a tricritical point (TCP) which separates a second-order phase
transition at high temperatures and low fields from a first-order phase transition at low temperatures
and high fields (Fig.1a). Although this picture has in principle been confirmed experimentally
\cite{stryjewski}, some ambiguities concerning the tricritical behaviour still remain. In particular,
the mean-field theory \cite{motizuki} (MFT) predicts splitting of the TCP into a critical endpoint
(CE) and a bicritical endpoint (BCE) (as shown in Fig.1b), if the ratio between total intrasublattice
ferromagnetic and total intersublattice antiferromagnetic couplings is sufficiently small. Such a way,
for example, if we increased the field at any temperature between $T_{\mathrm{CE}}$ and
$T_{\mathrm{BCE}}$, the system would first undergo a first-order transition from an antiferromagnetic
phase to a generally different antiferromagnetic phase, and then a second-order transition to a
paramagnetic phase. Since, as we know, the MFT neglects fluctuations, which could destroy this
``middle" phase, some more sophisticated methods have been employed to give an answer to the question
of whether the kind of phase diagram shown in Fig.1b can really exist or not. However, no definite
conclusions have been drawn yet. So far, a two-dimensional next-nearest neighbor model ($nnn-model$)
for Ising antiferromagnet has been studied by a Monte Carlo renormalization group
\cite{landau-swendsen} (MCRG) with no indications of the TCP splitting, however, transfer matrix
techniques \cite{herrmann} did not reproduce (possibly because of the limited strip widths which could
be studied) tricritical behaviour at very small ratios R. The same model has been also studied in
three dimensions, which is more interesting case from the experimental point of view as well as from
such a respect that fluctuations are smaller than in two dimensions and hence, the MFT predictions are
more likely to be correct.  Monte Carlo (MC) and MCRG methods applied to a variety of Hamiltonians for
both $meta-model$ and $nnn-model$ of an Ising antiferromagnet showed that the exponent behaviour was
consistent with the MFT, however, could not detect any change in the phase diagram itself
\cite{her-ras-lan}. However, more recent MC simulations on $nnn-model$ produced only tricritical
behaviour for $R \geq 0.05$ \cite{herrmann-landau}, i.e. the result contradicting to the MFT
predictions.
\newline
\indent Clear distinction in critical behaviour between two- and three-dimensional models was observed
in the antiferromagnetic Blume-Capel model, which, according to the MFT, should also show decomposition
of the TCP. While MC simulations produced only tricritical behaviour in two dimensions
\cite{kim-bla-car-wan}, they provided clear evidence of the decomposition into a CE and a BCE in a
three-dimensional model \cite{wang-kimel}. In light of the previous studies on Ising metamagnets, which
seemed to favor a non-decomposition of the TCP, quite perplexing results have been recently obtained by
both experimental \cite{kleemann1}-\cite{kleemann3} and Monte Carlo
\cite{selke-dasgupta}-\cite{pleimling-selke} studies on one of the typical Ising metamagnets -
$\mathrm{FeBr_{2}}$. Although they failed to confirm the decomposition of the TCP, they quite
convincingly showed the existence of anomalies of the magnetization and the specific heat, which could
be associated with the decomposition. These anomalies were attributed to the effectively weak
ferromagnetic intralayer interaction and the high interlayer coordination - features present in the real
compound.
\newline
\indent In this paper, we perform high precision MC simulations and provide convincing evidence of the
stability of the tricritical point in the three-dimensional $meta-model$ of an Ising metamagnet for $R
\geq 0.01$. Such a way, we show that merely weak, or even extremely weak, ferromagnetic intralayer
interaction should not cause decomposition of the TCP, as predicted by the MFT. Hence, in light of the
previous results obtained for $\mathrm{FeBr_{2}}$, the high interlayer coordination, which is present in
the real compound but absent in our model, seems to be of crucial importance in producing of such a
decomposition, if it really takes place. \vspace{6mm}
\newline
\noindent {\bf\Large{2.Model and simulation technique}}\vspace{3mm}
\newline
\indent The considered system is the spin-$\frac{1}{2}$ Ising metamagnet described by the Hamiltonian
\begin{equation}
H=-J_{1}\sum_{<i,j>}s_{i}s_{j}-J_{2}\sum_{<i,k>}s_{i} s_{k}-h\sum_{i}s_{i}\ ,
\end{equation}
where $s_{i}=\pm1$ is an Ising spin, $<i,j>$ and $<i,k>$ denote the sum over nearest neighbors in the
plane and in adjacent planes, respectively, and $h$ is an external magnetic field. We choose $J_1>0$
and $J_2<0$ so that each of the planes is ferromagnetic, but antiferromagnetically coupled to adjacent
planes.
\newline
\indent According to the MFT scenario, it is only the ratio $R = z_{1}J_{1}/z_{2}|J_{2}|$, where $z_{1}$
and $z_{2}$ are numbers of nearest neighbors of the site $i$ in the plane and adjacent planes,
respectively, which determines the phase diagram. While for $R>\frac{3}{5}$ it should look like the one
in Fig.1a, i.e. it should display a TCP with tricritical exponents $\alpha_{t}=\frac{1}{2},
\beta_{t}=\frac{1}{4}, \gamma_{t}=1, \lambda_{t}=\frac{1}{2}$, for $0<R<\frac{3}{5}$ the TCP is expected
to split up into a CE and a BCE, as shown in Fig.1b. Then, in the latter case, both the CE and the BCE
should probably keep usual three-dimensional critical exponents: $\alpha \approx 0.11, \beta \approx
0.32, \gamma \approx 1.24$. At $R=\frac{3}{5}$ the MFT predicts a four-order critical point with yet
different set of exponents.
\newline
\indent We have performed MC simulations on simple cubic lattice samples of linear sizes ranging from
$L$ = 16 to $L$= 40, assuming the periodic boundary condition throughout. We used an antiferromagnetic
initial spin configuration at low temperatures and the field not exceeding the critical value
$h_{c}(p)=pz_{2}|J_{2}|$, and a ferromagnetic one at high temperatures. As we moved in the
temperature-field $(T,h)$ space, we used the last spin configuration as an input for calculation at
the next point. Spin updating followed a Metropolis dynamics. Averages were calculated using at most
25,000 Monte Carlo steps per spin (MCS/s) after equilibrating over another 5,000 to 10,000 MCS/s.
Since we focused on the tricritical region, which for very small ratios $R$ lies at very low
temperatures, in order to prevent huge fluctuations at field-heating and field-cooling processes (the
path of measurement would be virtually parallel to the phase boundary), we only performed
$(h\uparrow)$+$(h\downarrow)$ loops, i.e. raised and lowered the field at fixed temperature and
measured :
\newline
the direct and staggered magnetizations $m$ and $m_{s}$, respectively
\begin{equation}
m=\langle M \rangle/N,\ {\mathrm{where}}\ M=|\sum_{i\in {\mathrm{A}}}s_{i}+\sum_{j\in
{\mathrm{B}}}s_{j}|\ ,
\end{equation}
\begin{equation}
m_{s}=\langle M_{s} \rangle/N,\ {\mathrm{where}}\ M_{s}=|\sum_{i\in {\mathrm{A}}}s_{i}-\sum_{j\in
{\mathrm{B}}}s_{j}|\ ,
\end{equation}
where $N$ is a total number of sites, and A, B denote sublattices made up of "spin-up" and "spin-down"
planes, respectively,
\newline
and the corresponding direct and staggered susceptibilities per site $\chi$ and $\chi_{s}$,
respectively
\begin{equation}
\chi=\frac{(\langle M^{2} \rangle - \langle M \rangle^{2})}{Nk_{B}T}\ ,
\end{equation}
\begin{equation}
\chi_{s}=\frac{(\langle M_{s}^{2} \rangle - \langle M_{s} \rangle^{2})}{Nk_{B}T}\ .
\end{equation}
These quantities were used to determine the nature as well as a location of the transition. The
simulations were performed on the vector supercomputer FUJITSU VPP700/56.\vspace{6mm}
\newline
\noindent {\bf\Large{3.Results and discussion}}\vspace{3mm}
\newline
\indent We studied phase transitions of the system for four different values of $R$: 0.5, 0.2, 0.05
and 0.01, with particular attention paid to the region of the expected TCP. Fig.2 depicts phase
boundaries in the vicinity of the TCPs for the respective values of $R$. At sufficiently low
temperatures, both the direct and staggered magnetizations displayed besides jumps also pronounced
hysteresis formed by the $(h\uparrow)$+$(h\downarrow)$ cycles - features which are characteristic for
a first-order transition. In Fig.2 the upper and lower branches in the low temperature region (blank
circles) are determined by the jumps of the $m_{s}$ curve in the $(h\uparrow)$ and $(h\downarrow)$
processes, respectively, and outline the region of the coexistence of the antiferromagnetic and
ferromagnetic phases. Increasing temperature makes the hysteresis shrink, and eventually disappear at
a point which gave us a first guess for the location of the TCP (arrows). Note, however, that if this
is not done for sufficiently large $L$, the above mentioned phenomena, accompanying first-order
transitions, could be suppressed by the finite-size effects, and hence, lattices of small $L$ are not
suitable for such a kind of investigation. Upon further increase in temperature, the $(h\uparrow)$ and
$(h\downarrow)$ magnetization curves collapse onto a single smooth curve, signifying a second-order
transition (filled circles). Its location was determined by the location of the staggered
susceptibility peak, extrapolated for $L$ brought to infinity. As we can see, none of the cases shown
in Fig.2 shows any indications of the TCP decomposition. On the other hand, according the MFT the
splitting should take place in each of the cases and in very noticeable scale (e.g., for $R=0.05$ the
separation between the CE and BCE temperatures is estimated to $50\%$). Note that, in contrast to the
results obtained for the three-dimensional $nnn-model$ \cite{herrmann-landau}, the varying value of
$R$ has virtually no influence on the slope of the transition in the tricritical region (however, it
is not possible to see it right away from Fig.2 because of different scales).
\newline
\indent At this stage, judging by the obtained results, we could draw a preliminary conclusion that
there is no decomposition of the TCP in this model, at least for $R \geq 0.01$. In order to confirm
this claim, we picked up the case of $R=0.01$ (it is the most likely candidate for the decomposition,
if there is any) and investigated it more closely. In particular, we ran additional simulations for
lattices of larger $L$ and finally extrapolated to $L \rightarrow \infty$. This data presented on a
fine scale allowed us to localize the TCP with fairly high accuracy to
$(k_{B}T_{t}/|J_{2}|,h_{t}/|J_{2}|)=(0.0150\pm0.0001,1.999138\pm0.000002)$. Then we examined behaviour
of some physical quantities as they approach the TCP. Fig.3 depicts a log-log plot of the direct and
staggered susceptibilities vs field near the TCP. While a good fit to a power-low behaviour can be
observed in a fairly narrow region quite near the TCP, a more distant region shows a deviation from
such a behaviour, probably as a consequence of logarithmic corrections predicted for tricritical point
in three dimensions \cite{wegner-riedel}. Nevertheless, slopes of the both lines seem to
asymptotically approach the magnitude close to 1 and $\frac{1}{2}$ for $\chi_{st}$ and $\chi$,
respectively. This means, however, that both the staggered and direct susceptibilities take on
exponents which are very close to the tricritical ones $\gamma_{t}=1$ and $\lambda_{t}=\frac{1}{2}$
rather than those which characterize an usual critical behaviour. These results just confirm the
previous conclusion about non-decomposition of the TCP and put it on firmer ground.
\newline
\indent In Fig.4 we depict both the tricritical temperature and the tricritical field vs $R$. While
the tricritical temperature shows a linear dependence over whole range of $R$, which for a certain
smaller range of $R$ was also observed in a diluted $nnn-model$ \cite{galam}, the tricritical field
displays a slight curvature. As expected, for $R\rightarrow 0$, the tricritical temperature moves to
zero, while the tricritical field approaches the exact value for the zero-temperature critical field
$h_{c}(0)/|J_{2}|=z_{2}(=2)$. \vspace{6mm}
\newline
\noindent {\bf\Large{4.Conclusions}}\vspace{3mm}
\newline
\indent We have investigated the possibility of the decomposition of the tricritical point in a
three-dimensional layered Ising metamagnet. Since the mean-field theory predicts such a decomposition
for small values of the ratio $R$, we mainly focused on that region. However, we observed only
tricritical behaviour with no signs of the decomposition for $R \geq 0.01$. In the case of the
smallest value of $R = 0.01$, where, according to the MFT, the TCP is most likely to decompose, we
managed to locate the TCP with a precision of approximately $1\%$ in temperature and $0.0001\%$ in the
field. Even for such a small $R$ the analysis of the critical exponents clearly showed tricritical
behaviour. Therefore, we conclude that it is very unlikely that the TCP decomposes for any value of
$R$, although some very small possibility of the decomposition for $R<0.01$ still remains. Hence,
recently found anomalies in $\mathrm{FeBr_{2}}$ (note that it has relatively high value of $R$
\cite{aru-kats-kat,selke-dasgupta}) make us believe that the high interlayer coordination (interlayer
superexchange paths present in the real material), supposedly causing the anomalies by inducing local
thermal excitations of the second antiferromagnetic phase for weak intralayer couplings $J_{1}$, could
play a key role in the possible TCP decomposition.

\begin{figure}[!t]
\subfigure{\includegraphics[scale=0.5]{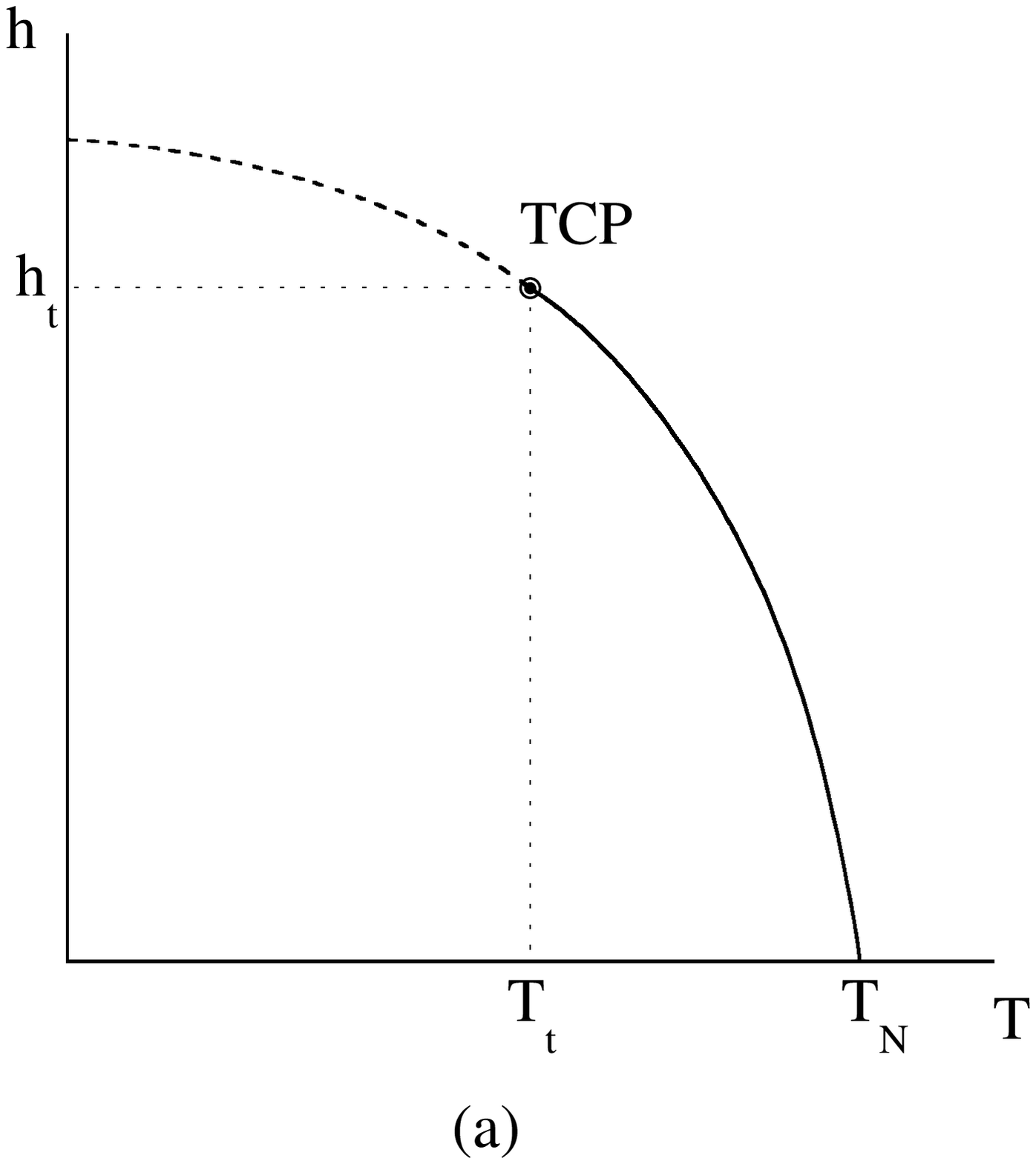}}
\subfigure{\includegraphics[scale=0.5]{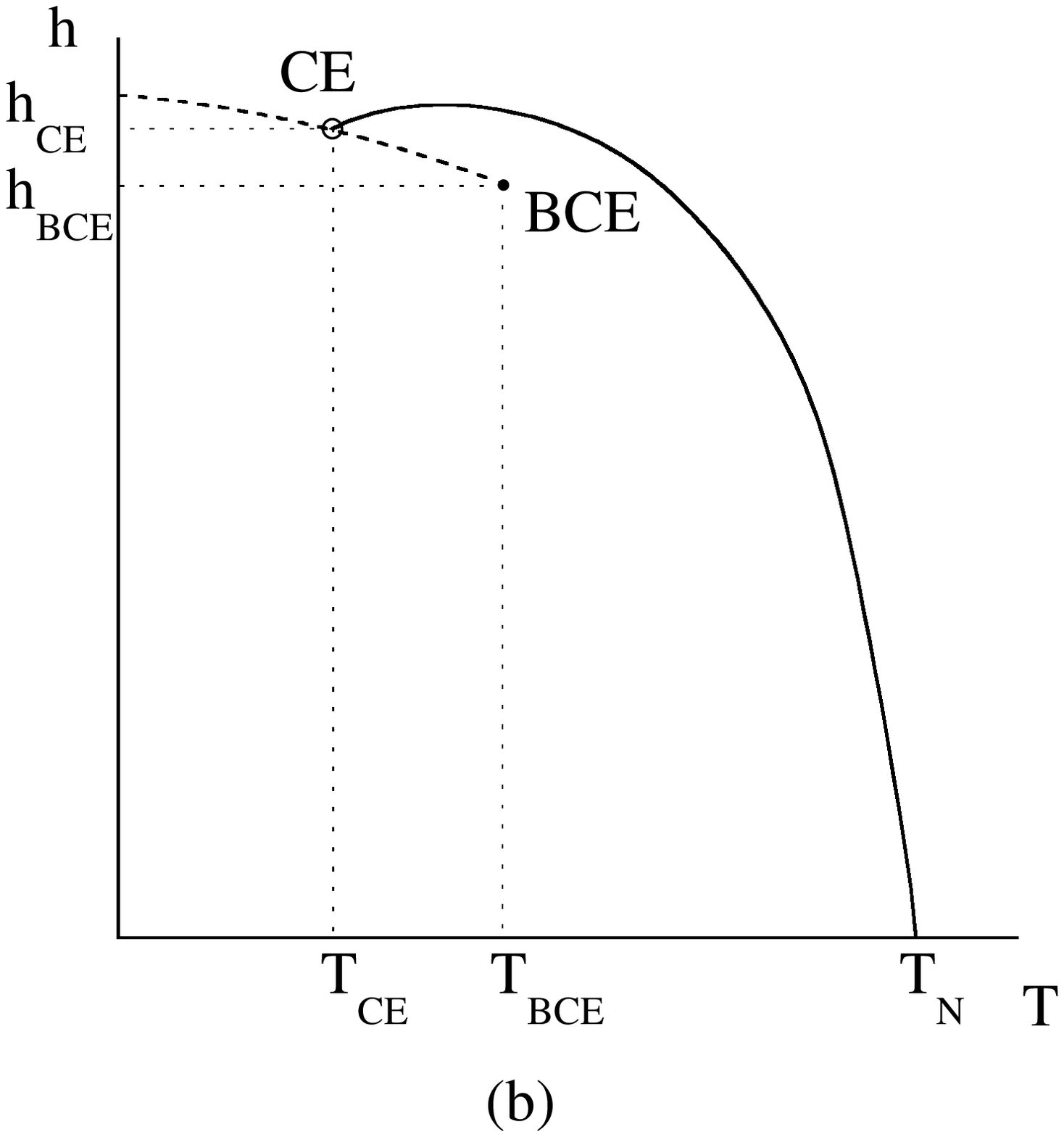}}
\caption{Schematic phase diagrams of a metamagnet displaying two different kinds
of behaviour: the second-order and first-order transition lines (solid and dashed, respectively)
either (a) meet at the tricritical point (TCP) or, (b) as predicted by the MFT, can end up by a
critical endpoint (CE) and a bicritical endpoint (BCE).}
\end{figure}

\begin{figure}[!t]
\subfigure{\includegraphics[scale=0.4]{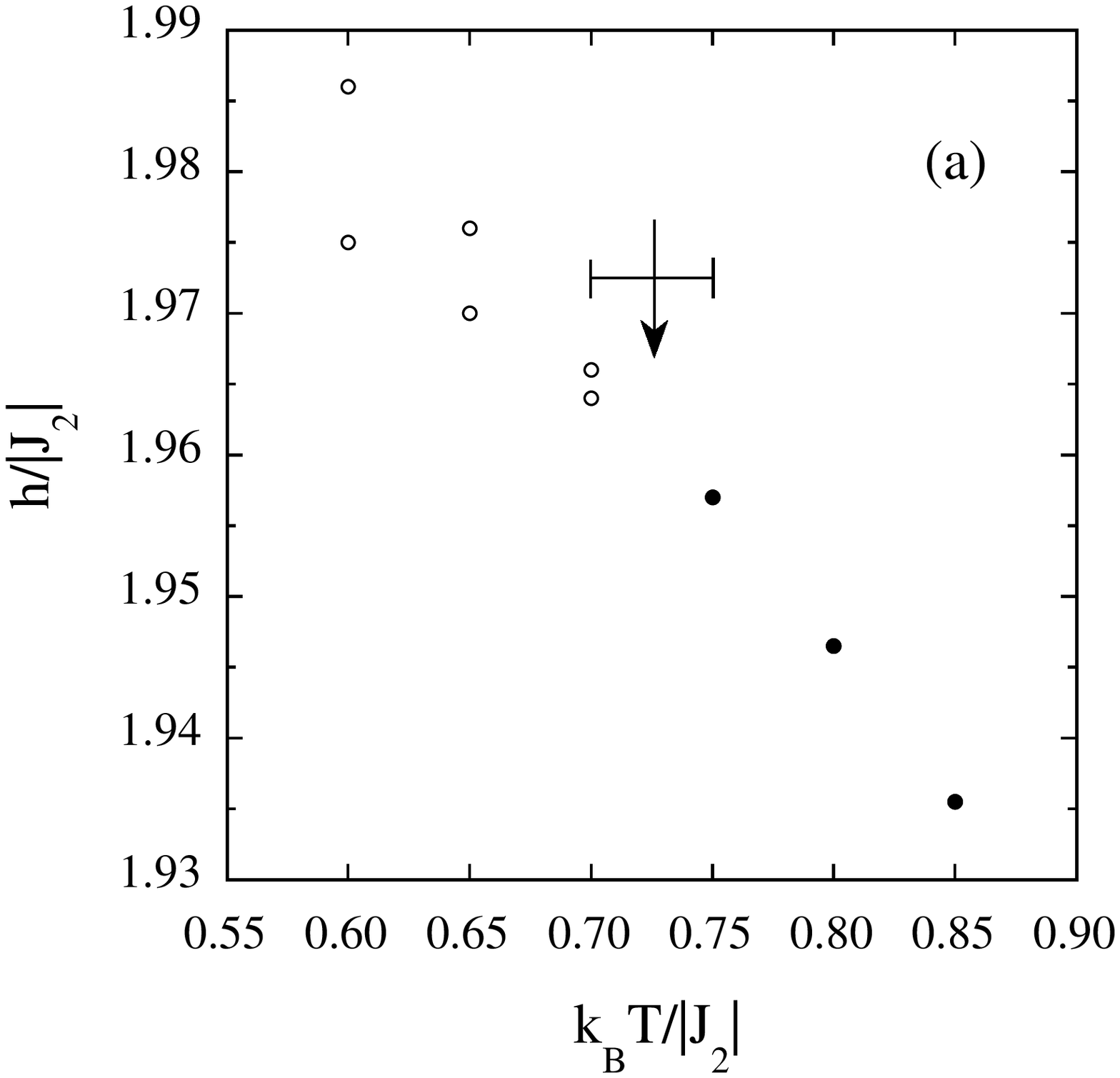}}
\subfigure{\includegraphics[scale=0.4]{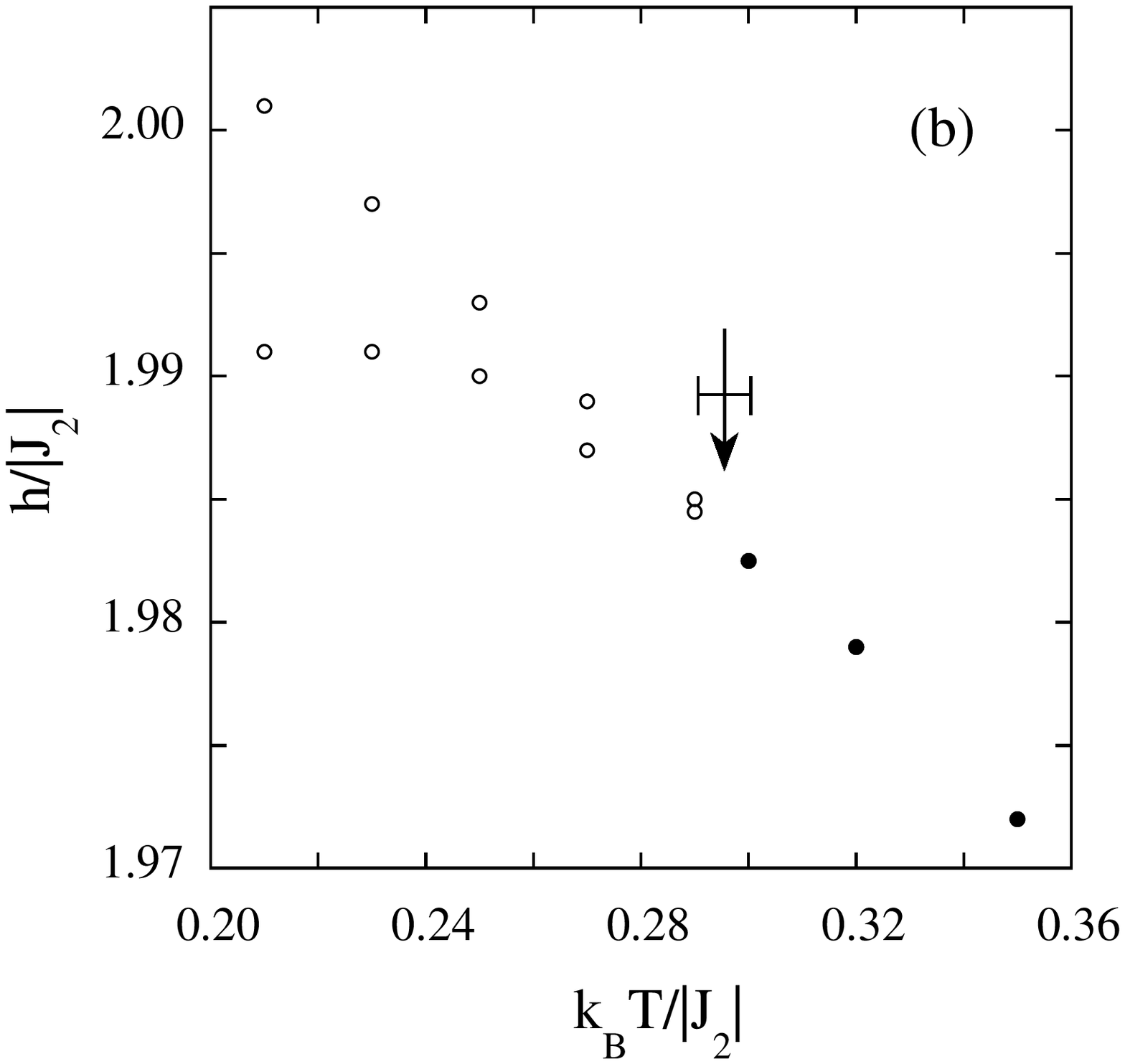}}
\subfigure{\includegraphics[scale=0.4]{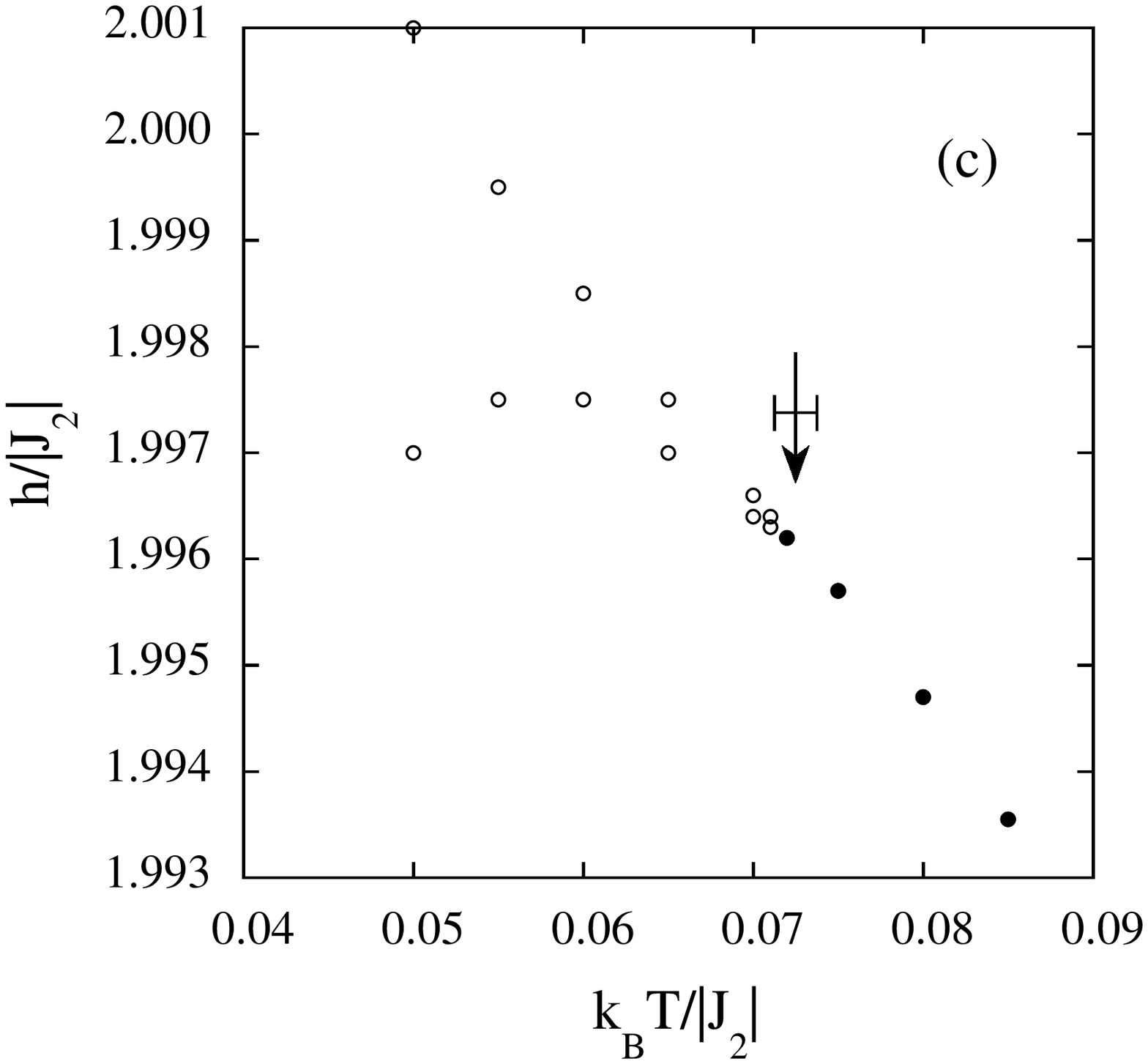}}
\subfigure{\includegraphics[scale=0.4]{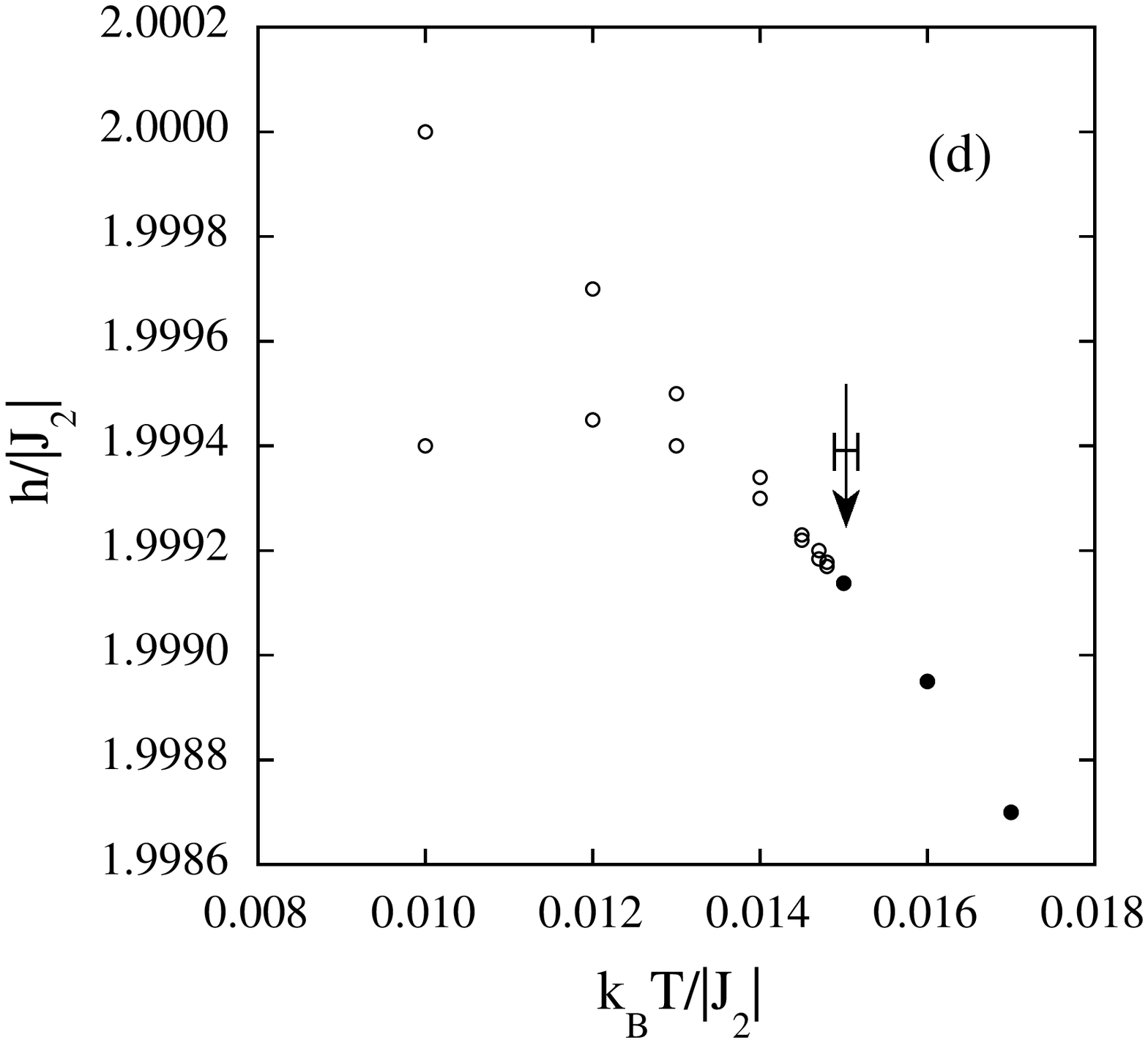}}
\caption{Phase diagrams near the TCP in the $T-h$ plane for four different values
of $R$: (a) 0.5, (b) 0.2, (c) 0.05 and (d) 0.01. The TCP (arrow) separates second-order transitions
(filled circles) from first-order transitions (blank circles).}
\end{figure}

\begin{figure}[!t]
\includegraphics[scale=0.5]{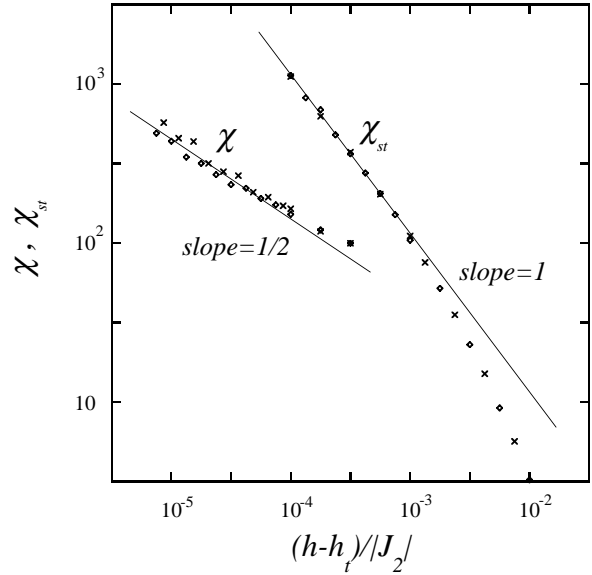}
\caption{Log-log plot of the direct and staggered susceptibilities $\chi$ and
$\chi_{st}$, respectively, as a function of $(h-h_{t})/|J_{2}|$ for $k_{B}T/|J_{2}|=0.0150$,
$h_{t}/|J_{2}|=1.999138$ and $R=0.01$. The crosses and the diamonds represent data for two different
lattice sizes $L=30$ and $L=40$, respectively, and the solid lines of slopes 1/2 and 1 (expected
tricritical exponents for $\chi$ and $\chi_{st}$, respectively) serve as guides to the eye.}
\end{figure}

\begin{figure}[!t]
\includegraphics[scale=0.5]{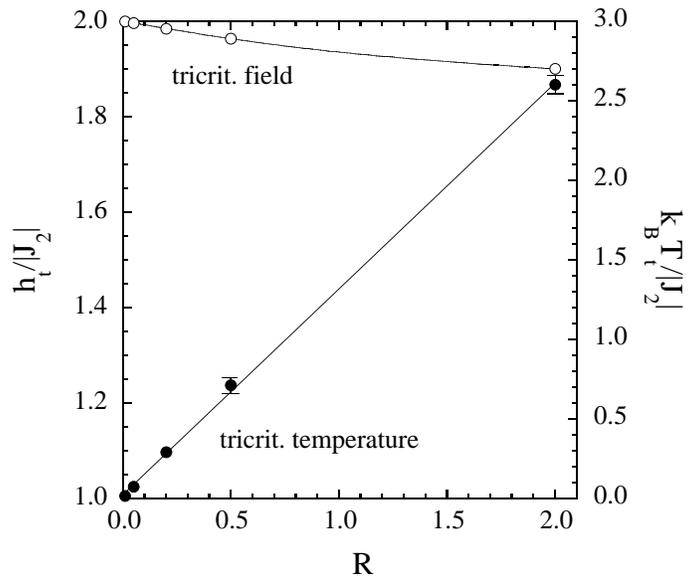}
\caption{Tricritical temperature $k_{B}T_{t}/|J_{2}|$ and tricritical field
$h_{t}/|J_{2}|$ as functions of the ratio $R$. Vertical bars indicate errors and are only shown when
they exceed the sizes of the symbols.}
\end{figure}

\end{document}